\documentclass[prl,reprint, nofootinbib, superscriptaddress]{revtex4-1}
\usepackage{graphicx, physics, amssymb, amsthm, amsmath, hyperref, xcolor, multirow, verbatim}

\begin{document}
\title{Experimental implementation of an efficient test of quantumness}

\author{Laura Lewis}
\email{llewis@caltech.edu}
\affiliation{Institute for Quantum Information and Matter and Department of Computing and Mathematical Sciences, California Institute of Technology, CA 91125, USA}
\affiliation{Division of Physics, Mathematics, and Astronomy, California Institute of Technology, CA 91125, USA}
\author{Daiwei Zhu}
\affiliation{Joint Quantum Institute, Departments of Physics and Electrical and Computer Engineering, University of Maryland, College Park, MD 20742, USA}
\affiliation{Joint Center for Quantum Information and Computer Science, NIST/University of Maryland, College Park, MD 20742, USA}
\affiliation{IonQ, Inc., College Park, MD  20740, USA}
\affiliation{Departments of Electrical and Computer Engineering, University of Maryland, College Park, MD 20742, USA}
\author{Alexandru Gheorghiu}
\affiliation{Institute for Theoretical Studies, ETH Z{\"u}rich, CH 8001, Switzerland}
\author{Crystal Noel}
\affiliation{Joint Quantum Institute, Departments of Physics and Electrical and Computer Engineering, University of Maryland, College Park, MD 20742, USA}
\affiliation{Duke Quantum Center and Department of Physics, Duke University, Durham, NC 27708, USA}
\affiliation{Department of Electrical and Computer Engineering, Duke University, Durham, NC 27708, USA}
\author{Or Katz}
\affiliation{Duke Quantum Center and Department of Physics, Duke University, Durham, NC 27708, USA}
\affiliation{Department of Electrical and Computer Engineering, Duke University, Durham, NC 27708, USA}
\author{Bahaa Harraz}
\affiliation{Joint Quantum Institute, Departments of Physics and Electrical and Computer Engineering, University of Maryland, College Park, MD 20742, USA}
\author{Qingfeng Wang}
\affiliation{Joint Quantum Institute, Departments of Physics and Electrical and Computer Engineering, University of Maryland, College Park, MD 20742, USA}
\affiliation{Joint Center for Quantum Information and Computer Science, NIST/University of Maryland, College Park, MD 20742, USA}
\affiliation{Chemical Physics Program and Institute for Physical Science and Technology, University of Maryland, College Park, MD 20742, USA}
\author{Andrew Risinger}
\affiliation{Joint Quantum Institute, Departments of Physics and Electrical and Computer Engineering, University of Maryland, College Park, MD 20742, USA}
\affiliation{Joint Center for Quantum Information and Computer Science, NIST/University of Maryland, College Park, MD 20742, USA}
\author{Lei Feng}
\affiliation{Joint Quantum Institute, Departments of Physics and Electrical and Computer Engineering, University of Maryland, College Park, MD 20742, USA}
\affiliation{Joint Center for Quantum Information and Computer Science, NIST/University of Maryland, College Park, MD 20742, USA}
\author{Debopriyo Biswas}
\affiliation{Joint Quantum Institute, Departments of Physics and Electrical and Computer Engineering, University of Maryland, College Park, MD 20742, USA}
\affiliation{Joint Center for Quantum Information and Computer Science, NIST/University of Maryland, College Park, MD 20742, USA}
\author{Laird Egan}
\affiliation{Joint Quantum Institute, Departments of Physics and Electrical and Computer Engineering, University of Maryland, College Park, MD 20742, USA}
\affiliation{Joint Center for Quantum Information and Computer Science, NIST/University of Maryland, College Park, MD 20742, USA}
\author{Thomas Vidick}
\affiliation{Institute for Quantum Information and Matter and Department of Computing and Mathematical Sciences, California Institute of Technology, CA 91125, USA}
\author{Marko Cetina}
\affiliation{Joint Quantum Institute, Departments of Physics and Electrical and Computer Engineering, University of Maryland, College Park, MD 20742, USA}
\affiliation{Duke Quantum Center and Department of Physics, Duke University, Durham, NC 27708, USA}
\author{Christopher Monroe}
\affiliation{Joint Quantum Institute, Departments of Physics and Electrical and Computer Engineering, University of Maryland, College Park, MD 20742, USA}
\affiliation{Joint Center for Quantum Information and Computer Science, NIST/University of Maryland, College Park, MD 20742, USA}
\affiliation{IonQ, Inc., College Park, MD  20740, USA}
\affiliation{Duke Quantum Center and Department of Physics, Duke University, Durham, NC 27708, USA}
\affiliation{Department of Electrical and Computer Engineering, Duke University, Durham, NC 27708, USA}

\date{\today}
\begin{abstract}
A test of quantumness is a protocol where a classical user issues challenges to a quantum device to determine if it exhibits non-classical behavior, under certain cryptographic assumptions.
Recent attempts to implement such tests on current quantum computers rely on either interactive challenges with efficient verification, or non-interactive challenges with inefficient (exponential time) verification. In this paper, we execute an efficient non-interactive test of quantumness on an ion-trap quantum computer. Our results significantly exceed the bound for a classical device's success.
\end{abstract}

\maketitle


As research in quantum theory continues to advance, experimentally testing the validity of the theory becomes of greater importance. In particular, a key question is whether quantum mechanics is falsifiable in the regime of high complexity arising from large entangled states~\cite{aharonov2012quantum}. This is exceptionally difficult to answer due to the exponential complexity in representing general quantum systems. Traditionally, one can test a physical theory by first predicting an outcome according to the theory and comparing with the experimental result. In quantum mechanics, such predictions can require exponential resources to obtain and therefore do not provide a feasible approach for validating the theory. 

An interesting alternative is to focus on verifying the quantum behavior of a quantum device via \emph{tests (or proofs) of quantumness}. In a proof of quantumness, a trusted classical user, known as the \emph{verifier}, wishes to determine if a quantum device, known as the \emph{prover}, indeed exhibits non-classical behavior. The verifier does so by issuing a series of challenges to be answered by the prover. The challenges are constructed so that a classical prover would be unable to answer them, unless it is able to efficiently solve hard cryptographic problems (such as factoring, or the Learning With Errors problem~\cite{regev2009lattices}).
On the other hand, the quantum prover is able to answer these challenges, without necessarily violating the intractability of the cryptographic tasks. Crucially, the verifier can efficiently check whether the challenges were answered correctly or not. 
This then serves as a test of quantum behavior under certain cryptographic assumptions.

Recently, several works have addressed the problem of constructing such cryptographic proofs of quantumness~\cite{brakerski2018cryptographic, kahanamoku2021classically, hirahara2021test,liu2021depth,zhu2021interactive}. All have in common the fact that they are \emph{interactive} protocols. 
In other words, the protocols work by having the verifier issue a challenge to the prover, the prover responds, the verifier issues another challenge and the process repeats. After a certain number of rounds, the verifier either accepts or rejects based on the prover's responses in all rounds. 
The main experimental challenge with such protocols is that the quantum prover must perform \emph{mid-circuit measurements} in order to correctly answer the verifier's challenges. The feasibility of doing this with near-term devices was recently demonstrated in~\cite{zhu2021interactive}. 

An alternative approach for certifying the quantumness of a device was introduced in~\cite{brakerski2020simpler}. This replaces the need for interaction with the use of a one-bit \emph{hash function}. The high-level idea is that because the hash function acts as a random function (or, more formally, as a \emph{random oracle}), in order to succeed in the verifier's new challenge involving the hash function, the prover must effectively have been able to answer both branches of the interactive version of the protocol. In a sense, the hash function accounts for both branches of the interactive protocol, eliminating the need for interaction. The idea of using hash functions to eliminate interaction originates in cryptography, where it is known as the Fiat-Shamir heuristic~\cite{fiat1986prove}.

This technique opens up more possibilities for efficient tests of quantum mechanics on near-term devices and contrasts the approaches used previously to certify quantum advantage~\cite{Arute2019, zhong2020quantum}. Those approaches are based on delegating a sampling task to the quantum device (such as random circuit sampling or boson sampling) and then checking the validity of the obtained samples using the \emph{linear cross-entropy benchmark (LXEB)}~\cite{aaronson2016complexity}. The major downside of this approach is that computing the LXEB takes exponential time, meaning that certifying quantum advantage in this way quickly becomes intractable~\cite{aaronson2016complexity, Arute2019, aaronson2019classical}. In addition, there are situations in which the LXEB can be ``classically spoofed'' (i.e. there is an efficient classical algorithm which can produce samples that are valid according to the LXEB)~\cite{barak2020spoofing, gao2021limitations}. 
The proof of quantumness of~\cite{brakerski2020simpler}, on the other hand, requires only polynomial runtime to perform the certification and is thus efficient. In addition, classically spoofing the results of a proof of quantumness is (provably) as hard as breaking the underlying cryptography. 
We note that a new non-interactive test of quantumness was recently introduced in~\cite{yamakawa2022verifiable}, which only relies on one of the two cryptographic assumptions required in~\cite{brakerski2020simpler} (the hash functions, see below). However, the protocol from~\cite{yamakawa2022verifiable} seems more computationally intensive for the quantum prover compared to the approach in~\cite{brakerski2020simpler}. For this reason, we only consider an experimental implementation of~\cite{brakerski2020simpler}.



In this work, we advance past the experimental work for the simpler learning with errors protocol in~\cite{zhu2021interactive} to eliminate interaction and implement the protocol of~\cite{brakerski2020simpler} on an ion-trap quantum computer using $11$ qubits out of its $13$ available qubits.
Our results are also complementary to the recent experimental work~\cite{stricker2022towards}, which implements a simpler version~\cite{carrasco2021theoretical} of Mahadev's interactive protocol for the classical verification of quantum computations~\cite{mahadev2018classical} (but does not experimentally implement the required interaction). 
In each of our experiments, the quantum device's success rate in answering the verifier's challenges significantly exceeds that of the best possible classical strategy. This therefore verifies our device's non-classical behavior and serves as a non-interactive proof of quantumness. We also comment on the possibility of scaling up this experiment to larger devices as a test of quantum mechanics.

The non-interactive protocol of~\cite{brakerski2020simpler} relies on two cryptographic primitives: \emph{trapdoor claw-free functions} (TCF) and \emph{hash functions}~\cite{GoldwasserMR84}.

A TCF, denoted $f$, is a 2-to-1 function. In other words, there exist exactly two preimages $x_0,x_1$ that map to the same image $w = f(x_0) = f(x_1)$. The pair $(x_0, x_1)$ is referred to as a \emph{claw}. The ``claw-free'' property of a TCF is that, given the description of $f$ (for instance, a circuit which evaluates $f$), it should be intractable to find a claw. In other words, no polynomial time classical (or quantum) algorithm can find a tuple $(x_0, x_1, w)$ such that $f(x_0) = f(x_1) = w$. Finally, the \emph{trapdoor} is a secret information that allows one to efficiently invert the function, recovering $x_0$ and $x_1$ from $w$. 

The TCF we consider in this paper is based on the \emph{learning with errors} (LWE) problem~\cite{regev2009lattices,regev2010learning}. In short, this problem is that of solving an \emph{approximate} system of linear equations over the integers modulo $q$, denoted $\mathbb{Z}_q$. Explicitly, given an $m \times n$ matrix $A \in \mathbb{Z}_q^{m\times n}$ with entries modulo $q$ and an $m$-dimensional vector $y = As + e \in \mathbb{Z}_q^m$ with entries modulo $q$, where $e$ is a vector with small entries, known as the error vector, the problem is to solve for $s \in \{0,1\}^n$. The entries in the error vector, $e$, are sampled from a discrete Gaussian distribution of small width (centered around $0$). The LWE problem is conjectured to be intractable for both classical and quantum computers (i.e. it cannot be solved in polynomial time), a fact known as the \emph{LWE assumption}~\cite{regev2010learning}. This assumed intractability forms the basis for defining a TCF. The specific TCF we consider here was also used in~\cite{zhu2021interactive, liu2021depth}. Starting from an LWE sample consisting of a matrix $A$ and vector $y = As + e$, the function is defined as
\begin{equation} \label{eqn:funrnd}
f(b,x) = \lfloor Ax + b y \rceil.
\end{equation}
Here, $b \in \{0,1\}$ is a single bit while $x \in \mathbb{Z}_q^n$ is a vector of dimension $n$ with entries modulo $q$. Additionally, $\lfloor \cdot \rceil$ denotes a rounding operation, which can be understood as taking the most significant bits of the entry being rounded (for more details, see the related learning with rounding problem~\cite{banerjee2012pseudorandom,alwen2013learning}). 
In this case, $\lfloor Ax + b y \rceil$ corresponds to simply taking the most significant bit of each component of the vector $Ax + by$. Notice that here the claw is determined by $f(0,x_0) = f(1,x_1)$ where $x_1 = x_0 - s$.

The second type of cryptographic function we consider is the hash function. Hash functions are a fundamental tool in cryptographic protocols and are usually modeled as \emph{random oracles}. An oracle function, $h: \{0,1\}^* \to \{0,1\}$, is a function for which one is not given an explicit description and instead queries it in a black box manner. Here $\{0,1\}^*$ denotes bitstrings of arbitrary length. A random oracle refers to the fact that the oracle function is chosen uniformly at random from the set of all functions (or rather, for each input length $n$, one chooses a random function from $\{0, 1\}^n$ to $\{0, 1\}$). As it is often difficult to prove the security of a protocol with respect to a concrete instantiation of a hash function, one instead proves security in the \emph{random oracle model}~\cite{bellare1993random}. This simply means that the hash function is modeled as a random function, which all parties in the protocol can evaluate. Classically, this means querying with some input $x$ and obtaining the output $h(x)$. In the quantum case, however, it is possible to query the random oracle in superposition~\cite{boneh2011random}. In other words, when performing a quantum query, the state $\sum_x \alpha_x \ket{x}\ket{y}$ is mapped to $\sum_x \alpha_x \ket{x}\ket{y \oplus h(x)}$.
Here, we restricted the output of the oracle to one bit, as this is the type of function used in the protocol of~\cite{brakerski2020simpler}. 

\begin{figure*}[htbp]
    \centering
    \includegraphics[width=17.2cm]{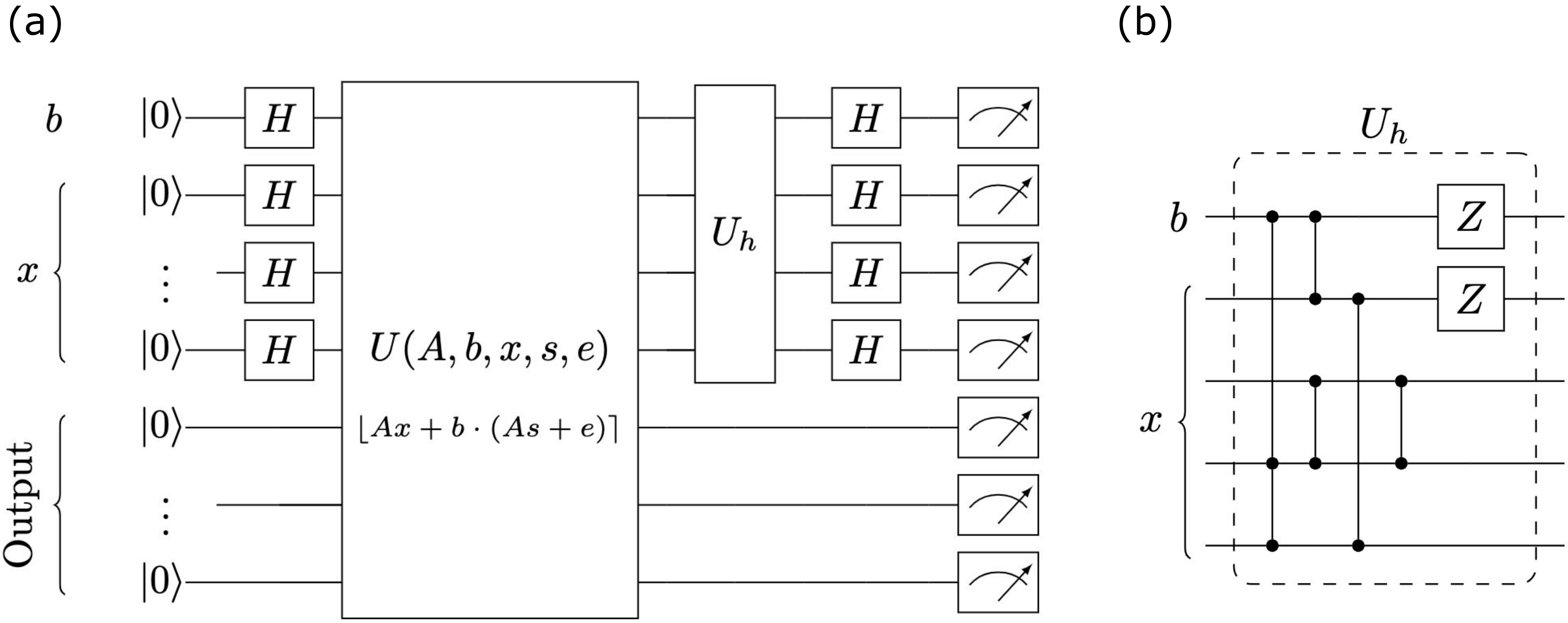}
    \caption{Circuit diagrams for prover's operations (a) and hash function used (b) in the protocol. The prover first evaluates the TCF $f$ and the hash function $H$ on a superposition of all possible inputs. In (a), $U(A,b,x,s,e)$ denotes the operations used to evaluate the TCF in Eq.~\ref{eqn:funrnd} and $U_h$ denotes the operations used to evaluate the cryptographic hash function in Eq.~\ref{eqn:hash}, which is illustrated explicitly in (b). Details about the implementation of $U(A,b,x,s,e)$ can be found in the supplementary information of \cite{zhu2021interactive}. The prover then measures the first two registers in the Hadamard basis and the third register in the standard basis.}
    \label{fig:circuit}
\end{figure*}

In the random oracle model and together with the LWE assumption, the protocol in~\cite{brakerski2020simpler} is a non-interactive test of quantum mechanics. When instantiating this protocol, we considered the TCF from Equation~\ref{eqn:funrnd} and a simple hash function represented as a low-degree polynomial. Ideally, one would use a hash function standardized by NIST, such as SHA-256 or SHA-3~\cite{national2015secure, dworkin2015sha}. However, those hash functions would require large numbers of qubits and gates in order to implement. For this reason, we propose using a small circuit, representing either a low-degree polynomial or a random (classical) circuit of short depth. This takes inspiration from the low-complexity hash functions introduced in~\cite{applebaum2017low} as well as the low-complexity one-way function of Goldreich~\cite{goldreich2011candidate}. Specifically, for our implementation we utilized the hash function

\begin{equation}
\label{eqn:hash}
H(b, x) = b + x_1 + b x_1 + x_2 x_3 + x_1 x_4 + x_2 x_3 + b x_3 x_4,
\end{equation}

\noindent where $x_i$ denotes the $i$th bit of the binary representation of $x$. It should be noted that in our implementation, $x = x_1 x_2 x_3 x_4$ is 4 bits long. A circuit diagram for this hash function, where the computation of the hash is performed in phase, is depicted in Figure~\ref{fig:circuit}(b).

With this background, we can now describe the protocol from~\cite{brakerski2020simpler} in more detail. A high-level circuit diagram depicting the prover's operations is displayed in Figure~\ref{fig:circuit}. 
Recall that the protocol is non-interactive, in the sense that it only consists of one challenge message from the verifier to the prover, followed by the prover's response. Additionally, one assumes that the hash function was chosen before the start of the protocol and known to both the verifier and the prover.

The protocol starts with the verifier generating an LWE instance $(A,y)$ (together with a trapdoor), that defines the TCF from Equation~\ref{eqn:funrnd} and sending the instance to the prover (while keeping the trapdoor secret). The prover is then required to evaluate the TCF $f$ and the hash function $H$ on a superposition of all possible inputs (consisting of the bit $b$ and the string $x$). The TCF is evaluated in the computational basis, while the hash function is evaluated \emph{in phase}.
In other words, the prover prepares the state
\begin{equation}
\sum_{b, x} (-1)^{H(b, x)} \ket{b, x} \ket{f(x)},
\end{equation}
suitably normalized. The prover then measures the third register, denoting the classical output as $w$, resulting in the state
\begin{equation}
\frac{1}{\sqrt{2}}(\ket{0,x_0} + (-1)^{H(0,x_0) + H(1,x_1)} \ket{1,x_1})\ket{w},
\end{equation}
where $(0, x_0)$ and $(1, x_1)$ are the two preimages of $w = f(0, x_0) = f(1, x_1)$.
Finally, the prover measures the qubits in the first two registers in the Hadamard basis. The prover's operations are depicted in Figure~\ref{fig:circuit}(a). Denoting the first bit in the measurement outcome as $m$ and the remaining bits as the string $d$, it can be shown that the following equation will be satisfied
\begin{equation} \label{eqn:equation}
d \cdot (x_0 \oplus x_1) = m \oplus H(0,x_0) \oplus H(1,x_1).
\end{equation}
When the verifier sent the LWE instance to the prover, the challenge for the prover was to produce a tuple $(w, m, d)$, such that Equation~\ref{eqn:equation} is satisfied. The quantum strategy of the prover, outlined here, does indeed produce such a tuple and this will be the prover's response to the verifier.
The verifier uses the trapdoor to invert $f$ on $w$, obtaining $(0, x_0)$ and $(1, x_1)$. With this, and the prover's response, the verifier checks Equation~\ref{eqn:equation}, accepting if it is satisfied and rejecting otherwise.
Note that while we presented the prover as performing its two measurements in sequence, the measurements can in fact be performed at the same time, as depicted in Figure~\ref{fig:circuit}(a). 

Let us provide some intuition for why a classical prover cannot succeed in the above protocol. The reason has to do with the intractability of finding a claw for the TCF and the fact that, classically, the random oracle (representing the hash function) can only be queried on a single input at a time (in contrast to the quantum case, where it is possible to query it on a superposition). We know that no efficient classical prover can produce a tuple $(w, x_0, x_1)$, with $f(0, x_0) = f(1, x_1) = w$. Of course, in the protocol, the prover is merely required to produce a valid equation in the preimages of $w$, which can in principle be easier than finding a claw. However, in this case the use of the hash function precludes this possibility. A classical prover cannot compute both $H(0, x_0)$ and $H(1, x_1)$, as this would require querying the oracle on both points, meaning that the prover had obtained a claw. This then means that at least one of $H(0, x_0)$, $H(1, x_1)$ will be random and so a classical prover's probability of finding a valid equation will be $1/2$. By simply repeating the protocol multiple times, the classical prover's probability of succeeding in all challenges becomes negligible.
The reason this argument fails for quantum provers is because quantum provers can query both the TCF and the random oracle in superposition. Indeed, this is precisely what is leveraged in the protocol in order to produce a valid equation. For the full proof of classical hardness, we refer the reader to~\cite{brakerski2020simpler}.


\begin{figure}[htbp]
    \centering
    \includegraphics[width=8.6cm]{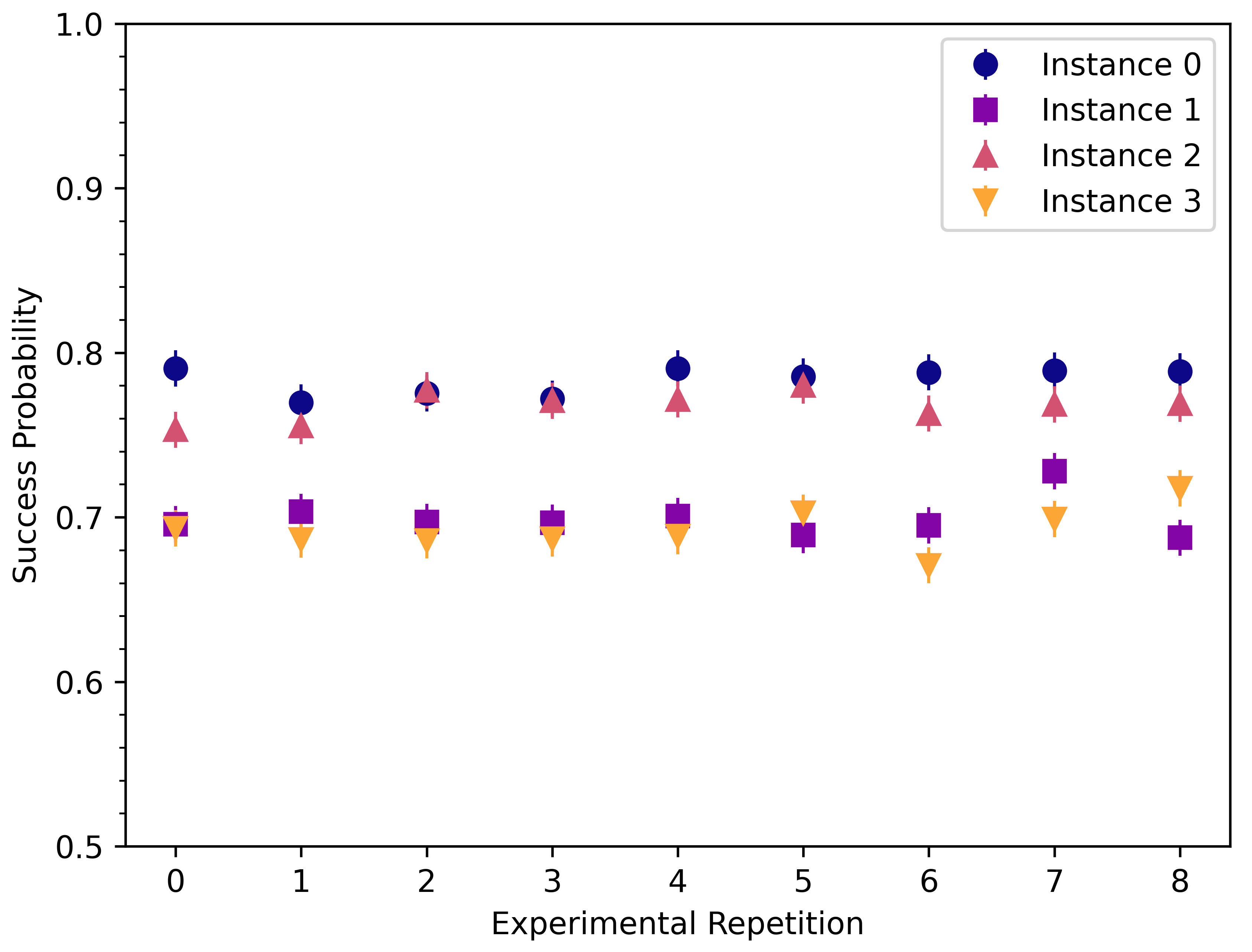}
    \caption{Results of the protocol on four different LWE instances from Table \ref{tab:lwe_instances}. The experiment is run $9$ times (experimental repetitions) for each instance with each repetition covering $2000$ executions of the experiment. The best possible success probability for a classical prover is $0.5$ while it is $1.0$ for an honest quantum prover. The error bars are using $\sigma = 1/(2\sqrt{N})$, where $N = 2000$ is the number of executions.}
    \label{fig:results}
\end{figure}

We implement the quantum prover's circuits (Figure \ref{fig:circuit}) using an ion-trap quantum computer \cite{zhu2021interactive} to test our protocol. The quantum computer consists of thirteen qubits made from a linear chain of fifteen $^{171}$Yb$^+$ ions that are laser cooled to near motional ground state. The system is capable of applying a universal gate set consists of arbitrary single-qubit rotations as well as two-qubit M{\o}lmer-S{\o}rensen gates\cite{Molmer99} on any target qubit. Individual-qubit readout is performed with high fidelity at the end of circuit operations. 
The results for implementing the protocols are displayed in Figure~\ref{fig:results}. Here, we ran the experiment for several different choices of the matrix $A$ and error vector $e$ in the LWE instance, detailed in Table~\ref{tab:lwe_instances} using 11 qubits.

\begin{table}[htbp]
    \centering
    \begin{tabular}{cccc}
       Instance & $A^\intercal$ & $e^\intercal$ & $(As + e)^\intercal$\\
       \hline
       0 & $\begin{pmatrix} 0 & 2 & 0 & 1\\2 & 0 & 1 & 2\end{pmatrix}$ & $\begin{pmatrix} 0 & 1 & 0 & 0 \end{pmatrix}$ & $\begin{pmatrix} 0 & 3 & 0 & 1 \end{pmatrix}$\\
       1 & $\begin{pmatrix} 0 & 2 & 3 & 2\\2 & 3 & 0 & 0 \end{pmatrix}$ & $\begin{pmatrix} 0 & 0 & 0 & 1\end{pmatrix}$ & $\begin{pmatrix} 0 & 2 & 3 & 3 \end{pmatrix}$\\
       2 & $\begin{pmatrix} 2 & 0 & 0 & 1\\0 & 3 & 2 & 1\end{pmatrix}$ & $\begin{pmatrix} 1  & 0 & 1 & 0\end{pmatrix}$ & $\begin{pmatrix} 3 & 0 & 1 & 1 \end{pmatrix}$\\
       3 & $\begin{pmatrix} 0 & 1 & 3 & 0\\3 & 0 & 0 & 2\end{pmatrix}$ & $\begin{pmatrix} 0 & 0 & 0 & 1\end{pmatrix}$ & $\begin{pmatrix} 0 & 1 & 3 & 1 \end{pmatrix}$\\
    \end{tabular}
    \caption{Details of the LWE instances. Note that the entries are transposed and for all instances we use $s^\intercal = \begin{pmatrix} 0 & 1\end{pmatrix}$.}
    \label{tab:lwe_instances}
\end{table}

Furthermore, we repeat the experiment $9$ times, with each repetition covering $2000$ executions of all four LWE instances in Table~\ref{tab:lwe_instances}. Hence, we obtain the success probabilities seen in Figure~\ref{fig:results} and Table~\ref{tab:results}. We note that Instances 0 and 2 perform better due to optimizations reducing the gate count for the implementation of the TCF based on those instances. In particular, we achieved a $42\%$ and $42.5\%$ decrease in gate count due to optimization for Instances 0 and 2, respectively. In contrast, the same optimization reduced the gate count of Instances 1 and 3 by only $25.3\%$ and $29.1\%$, respectively.

We know from~\cite{brakerski2020simpler} that a classical adversary can succeed in the verifier's challenge with probability at most $0.5$. Thus, we see that the results exceed this classical success probability for each LWE instance used. In particular, for each instance, we exceed this bound by at least $50\sigma$; see Table~\ref{tab:results}. Thus, our results significantly surpass the threshold for classical behavior, emphasizing our success in implementing a test of quantum mechanics experimentally, albeit on a small number of qubits. In particular, this confirms the quantum behavior of the device this was executed on given the cryptographic security of our TCF and hash function.

\begin{table}[htbp]
    \centering
    \[\begin{array}{c|c|c|c|c}
        \multicolumn{1}{c}{} & \multicolumn{4}{c}{\text{Instance}}\\
        & 0 & 1 & 2 & 3 \\
        \hline
        \text{Success Prob.} & 0.783 & 0.699 & 0.768 & 0.692 \\
        \text{Stat. Sign.} & 76.0\sigma & 53.3\sigma & 71.8\sigma & 51.6\sigma
    \end{array}\]
    \caption{Success probability and statistical significance for different LWE instances. The success probability is averaged over all $N = 18000$ trials, and $\sigma$ is computed using $\sigma = 1/(2\sqrt{N})$.}
    \label{tab:results}
\end{table}

We implemented an efficient non-interactive test of quantumness with an ion trap quantum computer and obtained results which exceed the threshold required for demonstrating non-classical behavior under certain cryptographic assumptions. Since our implementation used 11 qubits, this does not constitute a certification of quantum advantage and is instead certifying quantum mechanical behavior within the device. For a demonstration of quantum advantage, one would have to use a large enough instance of a claw-free function, for which classically ``breaking'' the underlying cryptographic task takes longer than the time it takes to run the experiment with a quantum device. The circuit complexity of implementing this function would be the dominating cost for the quantum prover's strategy. This is because, as we have shown with our implementation, the hash function can be implemented as some low-degree polynomial, or as a small random (classical) circuit. As such, the estimated numbers of qubits and circuit sizes for demonstrating quantum advantage with such a protocol are similar to those described in~\cite{zhu2021interactive, kahanamoku2021classically}, namely $\sim 10^3$ qubits and $\sim 10^5$ layers of depth. 
The main advantage of our protocol and implementation compared to those protocols is the fact that interaction is not required for performing the test of quantumness. Given this, as well as recent results aiming to reduce the costs of the quantum prover's implementation of the claw-free function~\cite{liu2021depth, hirahara2021test}, there are reasons to expect that these protocols could be used to certify quantum advantage on future generations of NISQ devices.

Finally, we note that the techniques used here can not only be applied to verifying quantum advantage, but also certifiable random number generation~\cite{brakerski2018cryptographic} and the classical verification of arbitrary quantum computation~\cite{mahadev2018classical}, for which non-interactive protocols have been achieved~\cite{brakerski2020simpler,chia2020classical,alagic2020non}. Although interactive protocols can also be utilized to accomplish these tasks and have been experimentally demonstrated~\cite{zhu2021interactive,stricker2022towards}, we emphasize that the non-interactive approach used here is simpler and thus more likely to be scalable. Thus by removing the additional barrier that interaction creates, these non-interactive protocols yield a promising path towards realizing tests of quantumness, randomness, and delegated computation on future quantum devices.


\section{Acknowledgements}
This work is supported by AFOSR YIP award number FA9550-16-1-0495, a Simons Foundation (828076, TV) grant, MURI Grant FA9550-18-1-0161, the NSF QLCI program (OMA-2016245), the IQIM, an NSF Physics Frontiers Center (NSF Grant PHY-1125565) with support of the Gordon and Betty Moore Foundation (GBMF-12500028), Dr.\ Max R{\"o}ssler, the Walter Haefner Foundation, the ETH Z{\"u}rich Foundation, a Caltech Summer Undegraduate Research Fellowship (SURF), the ARO through the IARPA LogiQ program, the NSF STAQ program, the U.S. Department of Energy Quantum Systems Accelerator (QSA) program, the AFOSR MURI on Scalable Certification of Quantum Computing Devices and Networks, the AFOSR MURI on Dissipation Engineering in Open Quantum Systems, and the ARO MURI on Modular Quantum Circuits.
\textbf{Competing Interests:} C.M. is Chief Scientist for IonQ, Inc. and has a personal financial interest in the company.

\bibliography{reference}

\end{document}